\documentclass[%
superscriptaddress,
twocolumn,
 amsmath,amssymb,
prb,
]{revtex4-2}

\usepackage{graphicx}
\usepackage{dcolumn}
\usepackage{bm}
\usepackage{xcolor}
\usepackage{hyperref}

\begin{document}


\title{Comparison of coherent phonon generation by electronic and ionic Raman scattering in LaAlO$_3$}

\author{Martin J. Neugebauer}
\affiliation{Institute for Quantum Electronics, Physics Department, ETH Zurich, CH-8093 Zurich, Switzerland}
\author{Dominik M. Juraschek}
\affiliation{Harvard John A. Paulson School of Engineering and Applied Sciences, Harvard University, Cambridge, MA 02138, USA}
\affiliation{Department of Materials, ETH Zurich, CH-8093 Zurich, Switzerland}
\author{Matteo Savoini}
\affiliation{Institute for Quantum Electronics, Physics Department, ETH Zurich, CH-8093 Zurich, Switzerland}
\author{Pascal Engeler}
\affiliation{Department of Materials, ETH Zurich, CH-8093 Zurich, Switzerland}
\author{Larissa Boie}
\affiliation{Institute for Quantum Electronics, Physics Department, ETH Zurich, CH-8093 Zurich, Switzerland}
\author{Elsa Abreu}
\affiliation{Institute for Quantum Electronics, Physics Department, ETH Zurich, CH-8093 Zurich, Switzerland}
\author{Nicola A. Spaldin}
\affiliation{Department of Materials, ETH Zurich, CH-8093 Zurich, Switzerland}
\author{Steven L. Johnson}
\affiliation{Institute for Quantum Electronics, Physics Department, ETH Zurich, CH-8093 Zurich, Switzerland}
\affiliation{SwissFEL, Paul Scherrer Institut, CH-5232 Villigen PSI, Switzerland}

\date{\today}

\begin{abstract}
In ionic Raman scattering, infrared-active phonons mediate a scattering process that results in the creation or destruction of a Raman-active phonon. This mechanism relies on nonlinear interactions between phonons and has in recent years been associated with a variety of emergent lattice-driven phenomena in complex transition-metal oxides, but the underlying mechanism is often obscured by the presence of multiple coupled order parameters in play. Here, we use time-resolved spectroscopy to compare coherent phonons generated by ionic Raman scattering with those created by more conventional electronic Raman scattering on the nonmagnetic and non-strongly-correlated wide band-gap insulator LaAlO$_3$. We find that the oscillatory amplitude of the low-frequency Raman-active $E_g$ mode exhibits a sharp peak when we tune our pump frequency into resonance with the high-frequency infrared-active $E_u$ mode, consistent with first-principles calculations. Our results suggest that ionic Raman scattering can strongly dominate electronic Raman scattering in wide band-gap insulating materials.  We also see evidence of competing scattering channels at fluences above 28~mJ/cm$^2$ that alter the measured amplitude of the coherent phonon response.
\end{abstract}

\maketitle

The interactions between electrons determine the structural and electronic properties of solids and depend strongly on the distance between the atomic nuclei. Ultrashort laser pulses are able to drive coherent vibrational motions of the atoms (coherent phonons) with large amplitudes, which in turn can lead to emergent phenomena, such as light-induced superconductivity \cite{Mankowsky2014,Mitrano2016,Liu2020}, ferroelectricity \cite{Nova2019,Li2019}, and phonon control of magnetic order \cite{Nova2017,Afanasiev2019,Disa2020,Juraschek2020_2}.
Conventionally, coherent phonons are created by ultrashort laser pulses in the visible spectral region which interact predominantly with interband electronic transitions. Electron-phonon coupling then results in an effective time-dependent force that drives coherent motion of the atoms~\cite{Merlin1997}. If the photon energy is far from a direct electronic resonance, the force arises from electronic Raman scattering (ERS) that can be described as a $E^2 Q_\mathrm{R}$ term in the second-order dipole coupling, where $E$ is the electric field and $Q_\mathrm{R}$ is the normal coordinate of the Raman-active phonon.  Although in principle one could achieve arbitrarily high coherent phonon amplitudes by increasing the peak electric field of the pump, in reality competing interactions such as multi-photon interband absorption lead to elevated temperatures and even irreversible damage to the material~\cite{Zukerstein2019}.

As an alternative mechanism, infrared-active phonons can replace electronic states in the scattering process. This so-called ionic Raman scattering (IRS) mechanism was predicted half a century ago \cite{Maradudin1970}. Only recently has it become possible to unequivocally observe the generation of coherent phonons via  IRS, largely because of new technologies to generate intense sources of radiation in the THz and mid-infrared frequency ranges~\cite{Forst2011}. IRS is based on nonlinear interactions between phonons due to anharmonicities in the potential energy landscape of the crystal lattice. An infrared-active phonon is initially coherently excited by a terahertz or mid-IR pulse through IR absorption and is subsequently scattered to a Raman-active phonon. The scattering process often involves a displacement of the potential energy surface of the Raman-active phonon toward a new quasi-equilibrium lattice structure~\cite{Mankowsky:2015,Mankowsky:2017,Mankowsky_2:2017}. The leading-order nonlinear phononic interaction in ionic Raman scattering can be described by a $Q_\textrm{IR}^2Q_\mathrm{R}$ in the potential energy of the lattice, where $Q_\textrm{IR}$ is the coordinate of an infrared active phonon driven by the pump~\cite{subedi:2014,subedi:2015,fechner:2016,Subedi2017,juraschek:2017,Gu2017,VonHoegen2018}.  This interaction is often invoked to explain a large variety of lattice-driven phenomena demonstrated in recent years. Due to the low photon energy of the pump relative to electronic transitions in insulating materials, IRS has proven to be highly selective \cite{Liu2020} and potentially less dissipative than ERS~\cite{Nicoletti2016}.

The goal of our study is to explicitly compare ERS and IRS without the interference from other order parameters that have been present in the vast majority of IRS demonstrations to date. In order to reduce possible extraneous influences from such interference, we study both mechanisms of coherent phonon generation in lanthanum aluminate (LaAlO$_3$), which we choose because it does not exhibit any magnetic, ferroelectric, or complex electronic order at any temperature. We measure an oscillatory response from coherent excitation of the $\sim$ 1~THz Raman-active $E_g$ mode in LaAlO$_3$ over a wide range of pump frequencies in the mid- and near-infrared.  Although the oscillations are measurable over the entire range of pump frequencies used, the amplitude of the response is maximized when the frequency of the laser is tuned into resonance with the high-frequency infrared-active $E_u$ mode at 20.6~THz. Our results are complementary to a recent study by Hortensius \textit{et al.}~\cite{Hortensius2020}, which reports similar measurements in LaAlO$_3$ with a focus on acoustic phonons.

\section{\label{sec:exp_setup}Experiment}

LaAlO\(_3\) is a wide band-gap insulator (\(E_{\rm gap}\) = 5.5 eV \cite{Lim2002}) that adopts the cubic perovskite structure with a low symmetry rhombohedral distortion (space group R$\bar{3}$c), characterized by alternating rotations of the AlO\(_6\) octahedra at temperatures below 813~K \cite{Hayward2005}. The samples used in our experiments are double-side polished bulk crystals with a thickness of 0.5~mm grown through the Czochralski method (MTI Corporation). The surface of these crystals is (1~0~0) using the pseudocubic unit cell.

A pump-probe scheme measures the response of the samples to mid-infrared excitation (see Appendix A). An amplified  Ti:Al\(_2\)O\(_3\) laser system consisting of an oscillator, regenerative amplifier and single-pass amplifiers creates femtosecond pulses of light (1~kHz, 800~nm, 110~fs). A pair of optical parametric amplifiers (OPAs) with a common white light seed generates independently tunable near-infrared signal pulses in a wavelength range between 1.2 and 1.6~\(\mu\)m.  These signal pulses in turn generate tunable, carrier-envelope phase-stable mid-infrared light via difference-frequency generation in GaSe~\cite{Junginger2010}.  An off-axis parabolic mirror focuses the generated mid-infrared light to a full width at half maximum (FWHM) spot diameter of 100~\(\mu\)m on the sample. For center frequencies below 33~THz, electro-optic sampling using another GaSe crystal of 30~\(\mu\)m thickness and a 15 fs probe pulse at 650~nm from a non-collinear OPA measures the electric field waveform.  For higher central frequencies a commercial Fourier transform interferometer characterizes the frequency content. The inset to Fig.~\ref{fig:fluence_dep}(a) shows the power spectral density from an example electro-optic sampling measurement of a pump pulse with center frequency near 20~THz. 
Fig.~\ref{fig:fluence_dep}(a) shows the corresponding time-domain electro-optic sampling measurement, where a Fourier filter is applied to cut off frequencies below 15~THz and above 22~THz to isolate the mid-infrared pulse component. The mid-infrared pulses have a FWHM duration (in intensity) of approximately 180~fs and were tuned between central frequencies of around 17 and 40~THz.  The maximum incident fluences depend strongly on the generated spectrum due to absorption in GaSe \cite{Mandal2008} and varied from several mJ/cm\(^2\) below 19~THz to more than 100~mJ/cm\(^2\) above 30~THz.

The probe pulses (650 nm, 15 fs) are transmitted through a small hole in the parabolic mirror and focused onto the sample at normal incidence to a FWHM spot diameter of 25~\(\mu\)m. The reflected pulses then propagate back through the parabolic mirror and are partially split into a balanced detection scheme that measures changes to the ellipticity $S$ (see Appendix A). Both pump and probe pulses are sent to the sample linearly polarized along [0~1~1] (pseudocubic). Since LaAlO\(_3\) is non-magnetic and non-absorbing at 650~nm \cite{Nelson2012}, the measured changes in ellipticity arise from the part of the probe beam that is transmitted into the sample and then reflected from the backside.


\section{\label{sec:results}Results}

\begin{figure}
\includegraphics[scale=0.7]{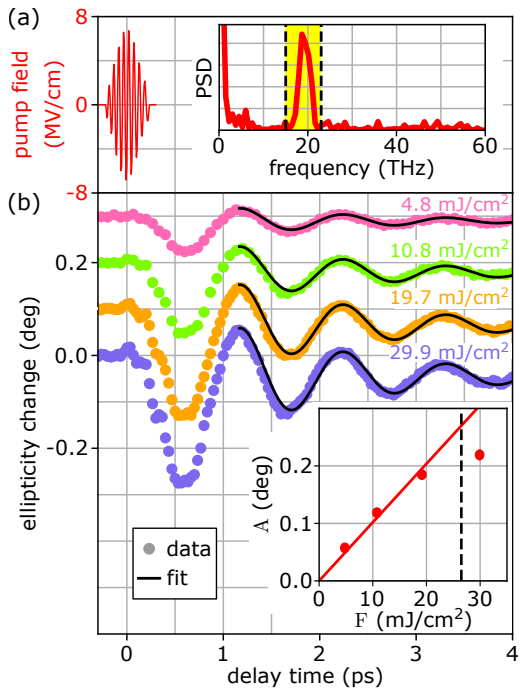}
\caption{\label{fig:fluence_dep} (a) Temporal evolution of the pump field according to Fourier-filtered electro-optic sampling (see text for details). The inset shows the power spectral density. The yellow-shaded area is the support of the Fourier filter for the temporal evolution. (b) Time-dependence of the ellipticity change \(S\) for various absorbed pump fluences. The data are shown together with fits (black lines, see text). The curves are offset for clarity. The inset shows how the amplitude \(A\) changes with fluence \(F\). Both panels share the delay time axis and the pump pulse displayed in (a) corresponds to an absorbed fluence of 29.9~mJ/cm\(^2\).}
\end{figure}

We acquired pump-probe data for a variety of pump fluences \(F\), center frequencies \(\nu_{\rm p}\), and positions on the sample. As an example, Fig.~\ref{fig:fluence_dep}(b) shows the measured changes in ellipticity \(S\) over delay time \(t\) when applying different pump fluences \(F\) for a center frequency of 19~THz. Here we use the absorbed fluence 
\begin{equation}
F = \frac{4 \ln 2 (1-R) U}{ \pi d^2}\label{eq:fluence}
\end{equation}
where $U$ is the pulse energy, and $d = 100\,\mu\textrm{m}$ is the FWHM of the spot on the sample. Here $R$ is an average of the reflectivity over a Gaussian-distributed range of frequencies with a center-frequency and FWHM matching  the pump pulse parameters, calculated using  complex permittivity values from the literature~\cite{Willett-Gies2014}. The pump pulse shown in panel (a) corresponds to the trace at \(F = 29.9\)~mJ/cm\(^2\). The data are characterized predominantly by an oscillation with a period of approximately 1~ps around a displaced value of \(S\). At pump-probe delay times corresponding to near-overlap (\(t = 0\)) there are additional, sharper features. To better isolate and quantify the oscillations, we fit the function
\begin{eqnarray}
S = A\sin\left(2\pi\nu t - \phi\right)e^{-t/\tau_\gamma} + B\left(1-e^{-t/\tau_{\rm R}}\right)\label{eq:model}
\end{eqnarray}
to all data sets for times \(t > 1.2\) ps, well beyond the times when the pump and probe pulses overlap. The first term describes the oscillations with amplitude \(A\), frequency \(\nu\), phase \(\phi\), and damping constant \(\tau_\gamma\). The second term represents a slower, non-coherent contribution with amplitude \(B\) that recovers with an exponential time constant \(\tau_{\rm R}\). In Fig.~\ref{fig:fluence_dep}(b) the model curves resulting from the fit  are shown as black lines. The parameters \(A\) and \(B\) are adjusted to best fit each data set individually, while \(\phi\) and \(\tau_{\rm R}\) are fit globally to each group of data sets for which the center frequency of the pump remained the same. The parameters \(\nu = 0.95\pm 0.04\)~THz and \(\tau_\gamma = 1.4\pm 0.3\)~ps are determined by a global fit to all data. These parameters agree well with previous measurements if we identify the oscillations as a coherent excitation of a degenerate pair of phonon modes with \(E_g\)-symmetry at room temperature~\cite{Scott1969}.  These \(E_g\)-modes are soft modes of the structural transition at 813~K.

\begin{figure*}
\begin{center}
\includegraphics[scale=0.7]{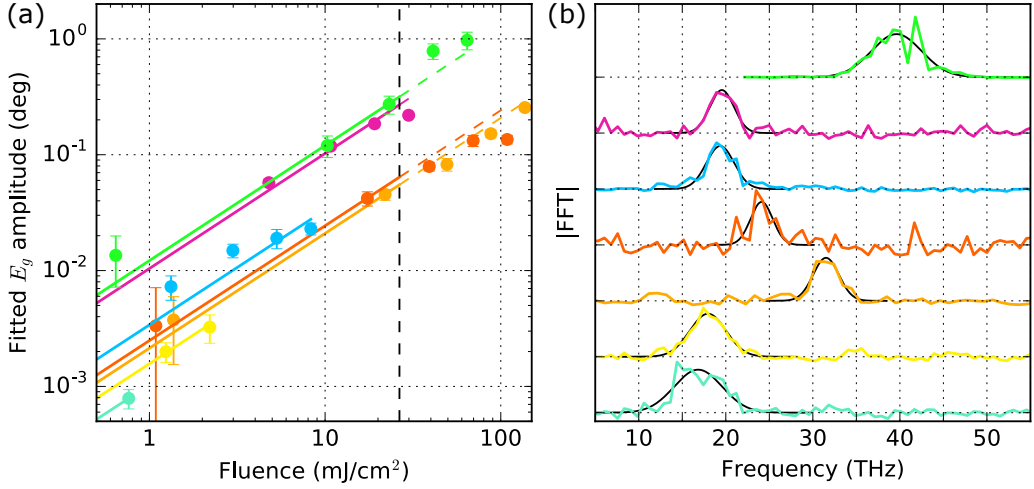}
\caption{\label{fig:scaling_various} Absorbed fluence dependence of the oscillations at different pump frequencies.  (a) Scaling of the \(E_g\) mode amplitude \(A\) with the absorbed fluence calculated using Eq.~\ref{eq:fluence} for various pump spectra including linear fits up to the fluence \(F_{\rm c}\) (dashed vertical line). (b) Different normalized pump spectra featuring Gaussian fits (black lines). The colors of the spectra in (b) correspond to the data points and fits shown in (a).}
\end{center}
\end{figure*}

The inset of Fig.~\ref{fig:fluence_dep}(b) shows the scaling of the amplitude \(A\) with the fluence \(F\) for the data shown on the main panel. While the behavior of \(A\) for the lower three fluences is consistent with a linear relationship \(A = a F\) (fit shown by the red line), the highest fluence at 29.9 mJ/cm$^2$ deviates from this proportionality. Fig.~\ref{fig:scaling_various} shows the corresponding fluence dependence on a log-log scale for a variety of pump frequencies.
Deviations from linearity at these high fluences are also evident for these measurements. For fluences less than $F_c$ = 28 mJ/cm$^2$ we fit these data to \(A = a F\) to extract a pump-frequency-dependent parameter $a$ that gives a quantitative measure of the sensitivity of the coherent phonon response to the pump at a given absorbed fluence.

\begin{figure*}
\includegraphics[scale=0.275]{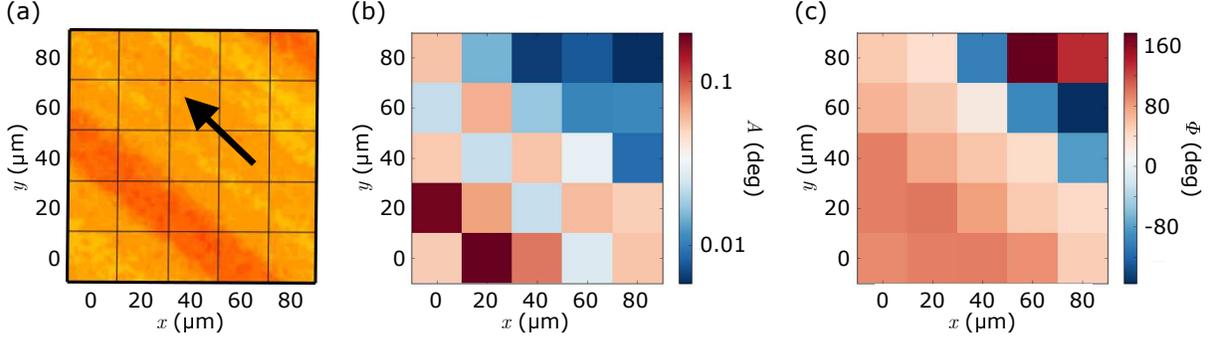}
\caption{\label{fig:grid} Spatially-resolved maps of a selected area of the sample. (a) optical microscope image showing stripes oriented along the [010] direction (indicated by the arrow), (b) amplitude \(A\) on logarithmic color scale, and (c) phase \(\phi\) on a linear color scale, extracted from fits of transient data taken at each point to Eq.~\ref{eq:model}.}
\end{figure*}

The rhombohedral axis of the low-symmetry, room-temperature structure of LaAlO$_3$ can lie along any of the body diagonals of the high temperature cubic cell, giving rise to distinct structural domains, which can lead to twinning~\cite{Yao1992, Hayward2002}. The different relative orientation of these domains results in a small amount of optical contrast.  Fig.~\ref{fig:grid}(a) shows an image of a 100~\(\mu\)m \(\times\) 100~\(\mu\)m area of the sample taken using an optical microscope. Here \(y\) corresponds to [0~1~1] and \(x\) to [0~-1~1] of the pseudocubic cell. Stripes of varying optical contrast  along [0~1~0] with a width on the order of 10~\(\mu\)m are evident, suggesting the presence of lamella-like domains. We investigated how these domains affect the oscillatory response of the system and mapped the 100~\(\mu\)m \(\times\) 100~\(\mu\)m area by scanning the position of the pump and probe focus positions in steps of 20~\(\mu\)m. The applied pump pulses had a central frequency of 23.8~THz and a FWHM bandwidth of 9.1~THz. For each position we fit \(S\) to the resulting traces to extract \(A\) and \(\phi\) as a function of the spot location. Fig.~\ref{fig:grid}(b) shows the respective map of \(A\) on a logarithmic color scale, while panel (c) shows the map of \(\phi\) on a linear scale. Over the mapped area \(A\) varies by a factor of more than 30, while \(\phi\) varies between -100\(^\circ\) and 180\(^\circ\). Stripes along [010] are evident in both parameter maps, suggesting a strong influence of the domain structure on the measured oscillations.

\section{\label{sec:discussion}Discussion}



While the parameter \(a\) that relates the absorbed fluence to the oscillation amplitude in principle reflects the relative sensitivity of the \(E_g\) mode to different excitation frequencies,  the oscillation amplitude of the coordinate \(Q_{\rm R}\) also depends on other quantities that change strongly with the pump frequency. Both the pump and probe pulses interact with the material with a strong dependence on both time and distance from the surface.  These spatial and temporal dynamics of the pump pulse vary considerably with frequency, necessitating a correction to the value of $a$  in order to genuinely compare the dependence of the \(Q_\textrm{R}\) on the absorbed pump fluence for different pump wavelengths.

In order to estimate the required correction factor, we assume that the time- and space-dependent coordinate of the Raman active mode is given by
\begin{equation}
Q_R(z,t) = \int\limits_{-\infty}^\infty dt^\prime \left|E_\textrm{pump}(z,t^\prime)\right|^2 G_{\nu_p}(t-t^\prime)
\end{equation}
where 
\begin{align}
E_\textrm{pump}(z,t) = & E_0 \tau \sqrt{\frac{\pi}{2 \ln 2}} \int\limits_{-\infty}^\infty d\nu \frac{2}{1+n(\nu)} \nonumber\\
& \times \mathrm{exp}\left[-\frac{\pi^2\tau^2(\nu-\nu_p)^2}{2\ln 2}\right] \nonumber\\
& \times \mathrm{exp}\left[i 2 \pi \nu \left[t - n(\nu) \frac{z}{c}\right]\right]
\end{align}
is the complex-valued electric field inside the crystal from a Gaussian-profile pump with peak incident amplitude $E_0$ at a depth $z$ and a time $t$, $n(\nu)$ is the complex-valued index of refraction at the frequency $\nu$, $\tau$ is the FWHM pulse duration, and $G_{\nu_p}$ is an impulse response function that depends on the pump frequency $\nu_p$.  Assuming that the probe pulse duration is small compared to the dynamics of \(Q_R\), the probe ellipticity change is
\begin{equation}
S(t) \propto \int\limits_0^D dz Q_R\left(z,t+\frac{z}{v_g}\right)
\end{equation}
where \(D\) is the thickness of the sample and \(v_g\) is the group velocity of the probe.  Here we are concerned only with the component of $S$ that oscillates at the phonon frequency \(\nu = 0.95\,\textrm{THz}\). We then have
\begin{equation}\label{eq:lin_fit}
A \propto |\tilde{G}_{\nu_p}(\nu) f(\nu)| F_\textrm{abs}
\end{equation}
as a  relation between the observed oscillation amplitude $A$ and the Fourier transform \(\tilde{G}_{\nu_p}\) of the impulse response function, where
\begin{align}
f(\nu) = & \frac{1}{1-R} \int\limits_0^D dz e^{i 2 \pi  \nu z/v_g} \\
& \times \int\limits_{-\infty}^\infty dt \left|\frac{E_\textrm{pump}(z,t)}{E_0}\right|^2  e^{-i2 \pi \nu t} \label{eq:f}
\end{align}
is a correction factor. We then define a new quantity 
\begin{equation}
a_\textrm{corr} \equiv a / |f(\nu)|\label{eq:acorr} 
\end{equation}
that depends on $\nu_p$ only via $\tilde{G}_{\nu_p}(\nu)$.  We  calculate $a_\textrm{corr}$ from Eqs.~\ref{eq:f} and \ref{eq:acorr} for our experiment taking optical constants from Ref.~\onlinecite{Willett-Gies2014} and assuming Gaussian pump pulses with a 100 fs FWHM.  

\begin{figure}
\includegraphics[scale=0.825]{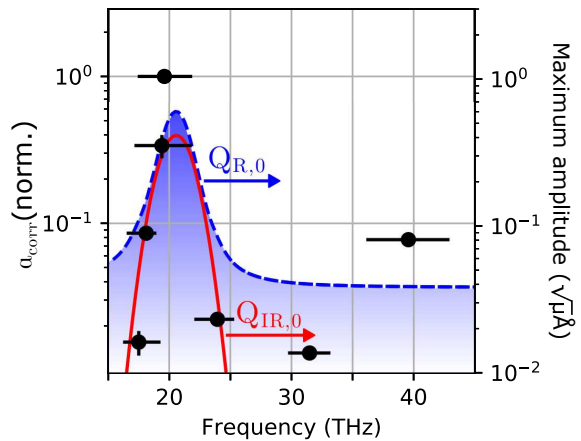}
\caption{\label{fig:resonance} Fluence-corrected scalings, \(a_{\rm corr}(\nu_{\rm p})\) (points, left axis), of the \(E_g\)-phonon amplitude and calculation values of the phonon-amplitude responses (lines, right axis) on logarithmic scales. For \(a_{\rm corr}\), the \(\nu_{\rm p}\) values are set according to the center of mass of the absorbed excitation spectrum, of which the horizontal errorbars indicate the FWHM. We show  the calculated frequency-dependent amplitudes of both the Raman-active \(E_g\)-mode \(Q_{\rm R, 0}\) and the infrared-active \(E_u\)-mode \(Q_{\rm IR, 0}\).}
\end{figure}

Values of $a_\textrm{corr}$ as a function of pump frequency for one particular region of the sample are shown in Fig.~\ref{fig:resonance}.  
There is a significant enhancement of $a_\textrm{corr}$ at frequencies near 20~THz.  We also observe a smaller enhancement when the pump is tuned to near 40~THz.  Due to the domain structure of the sample we do not attempt to compare the absolute magnitude of the response measured at different regions, although the relative response for a subset of pump frequencies and fluences was reproduced at several different sites.

Two mechanisms could lead to coherent oscillations of the \(E_g\) mode. The first is electronic Raman scattering (ERS). The two most prominent manifestations of ERS are electronically-driven displacive excitation and impulsive stimulated Raman scattering \cite{Merlin1997,Stevens2002}. In the former casr of displacive excitation, electrons are excited to higher bands in metals or semiconductors, which displaces the effective potential energy surface of the lattice.  This causes the ions to move coherently, oscillating about a new equilibrium position \cite{Zeiger1992}. In the latter case of impulsive Raman scattering, the energy of the photons is not large enough to persistently excite valence electrons into higher bands.  The resulting effective force on the ions exists only during the pump pulse interaction~\cite{Stevens2002,Glerean2019}. For the measurements reported here the photon energy is very far from an electric-dipole allowed transition, and so we are in the impulsive Raman scattering regime. It has previously been shown that at higher pump frequencies this mechanism can excite measurable $E_g$-mode oscillations in LaAlO$_3$~\cite{Liu1995}. In the impulsive limit we would expect the phase of the oscillations $\phi$ to be close to zero, and the magnitude of the oscillations to be nearly frequency independent, since the photon energy is very small compared to the band gap of 5.5~eV. The non-zero values of $\phi$ and the strong pump-frequency dependence of $a_\textrm{corr}$ both indicate that ERS alone is not sufficient to explain our data. A second possible mechanism is ionic Raman scattering arising from coupling via potential energy terms proportional to \(Q_{\rm IR}^2 Q_{\rm R}\), where $Q_\textrm{IR}$ is the normal coordinate of an infrared-active phonon mode at 20.6 THz~\cite{Scott1969}. 

To quantify the relative contributions of the two mechanisms, we perform simulations of both interactions using a combination of density-functional theory calculations and phenomenological modeling (see Appendix B). For the simulations we assume a constant envelope FWHM duration of 250~fs and peak electric fields of 12~MV/cm with varying central frequency \(\nu_{\rm p}\). From these we compute the maximum amplitudes of both the \(E_u\) mode \(Q_{\rm IR, 0}\) and the \(E_g\) mode \(Q_{\rm R, 0}\). \(Q_{\rm IR, 0}\) was calculated through the coupling between the pump field and the mode effective charge. For \(Q_{\rm R, 0}\) both IRS and ERS are taken into account. The simulations assume that the pump-field polarization is aligned  orthogonal to the c-axis of the crystal, and we compute the amplitude of the $E_g$ mode that couples most strongly to the pump. This is a simplification of the experimental conditions, where the pump and probe polarizations depend strongly on the domain orientation and is a priori not known because of the complex twinning.  This makes the calculated amplitude of the phonon response difficult to directly compare with experiment, but should give approximate indications of the relative efficiency of phonon generation as a function of pump frequency on the same location on the sample.  The dependence of the simulated responses on \(\nu_{\rm p}\) are shown in Fig.~\ref{fig:resonance} with the relevant scale on the right-hand axis. \(Q_{\rm IR, 0}\) is shown as a solid red and \(Q_{\rm R, 0}\) as a dashed blue line, reproducing the resonant behavior of the coherent phonon amplitude for pump frequencies near $Q_\textrm{IR}$. The strong enhancement of $a_\textrm{corr}$ in the vicinity of the $E_u$ mode at 20.6~THz and the agreement with the predictions of the simulations provide strong evidence that IRS is the dominant mechanism for the coherent driving of the 1.1 THz $E_g$ mode for pump frequencies ranging from 16-25~THz. Outside this range ERS plays a more significant role, and is largely characterized by a nearly frequency independent impulsive excitation.  Note that this does not explain the enhancement in $a_\textrm{corr}$ at 40 THz.  Measurements on other locations on the sample (using fewer pump frequencies) reproduced the enhancement near 20 THz but not the one at 40 THz, and so we do not consider the 40 THz enhancement further. We suspect that this apparent enhancement may be the result of a small change in the probe spot location in the sample that leads to different domain contributions that may affect the assumptions used to estimate $a_\textrm{corr}$.  

We also note that, even discounting the 40 THz pump data, there is a discrepancy between the experimental results and the simulations regarding the relative magnitudes of the IRS and ERS signals.  While the simulations show a difference of about a factor of 20 between the amplitudes of IRS and ERS driven coherent phonons, the experiment shows an enhancement of nearly 100 between the peak response at 20 THz pump and the response at pump frequencies near 15~THz or 30~THz.  Some of this discrepancy may arise from the fact that the polarization of the pump in the experiment differs from that assumed in the simulations, since components of the pump polarization orthogonal to the c axis can affect the efficiency of ERS (see Appendix B).

One striking feature of our data is the observation that for fluences above 28 mJ/cm$^2$ the amplitude of oscillations $A$ is no longer proportional to the absorbed fluence. This result deviates from the expectation of our simple models of ERS and IRS.  At most pump wavelengths the measured result for higher fluences is  sublinear, although at 40~THz pump frequencies we observed slightly superlinear behavior at these fluences. The value of the pump fluence at which deviations from linear behavior become measurable is relatively frequency independent and suggests that it is not related to a nonlinear phononic interaction, but instead to an optical phenomenon. The fact that our photon energies are less than $5$\% of the band gap suggests that these effects may be due to impurities or other defects that create a small number of carriers that then interact with the intense pump field. These effects could modify the magnitude of $Q_R$ and/or the phase matching conditions that allow us to detect the coherent phonon via ellipticity changes.

\section{\label{sec:conclusion}Conclusion}





Our data show clear evidence for identifying IRS as the dominant mechanism for coherent excitation of soft $E_g$ modes in LaAlO$_3$ when pumped with frequencies between 18 and 23 THz, with ERS playing a significant role at frequencies between 25 and 40 THz. Future experiments using ultrafast x-ray diffraction on detwinned samples could be used to further test the quantitative correspondence between the predictions of density-functional-theory-based simulations and experiment. Interestingly, despite the fact that the photon energies of the pump pulses were an extremely small fraction of the optical band gap, we observed nonlinear responses at fluences above 28~mJ/cm$^2$. These fluences are considerably less 
than those used in recent experiments where IRS was considered to be the only mechanism of interaction due to the band-gap mismatch~\cite{Mankowsky2017,VonHoegen2018}. In such experiments, it may therefore become necessary to consider additional interactions in this excitation range, even for wide band-gap materials.

Finally, we note that comparisons of electronic and ionic excitation mechanisms have recently been discussed both theoretically \cite{Maehrlein2017,Juraschek2018a} and experimentally \cite{Maehrlein2017,Melnikov2018,Johnson2019,Knighton2019,Melnikov2020} for the sum-frequency counterparts of ERS and IRS. These sum-frequency mechanisms are fundamentally two-photon and two-phonon absorption processes (compared to the difference-frequency Raman scattering processes here) and follow the same theoretical formalism as we use in this study, however at different spectral ranges.







\appendix

\section{Experimental details}

\begin{figure*}
\includegraphics[scale=0.475]{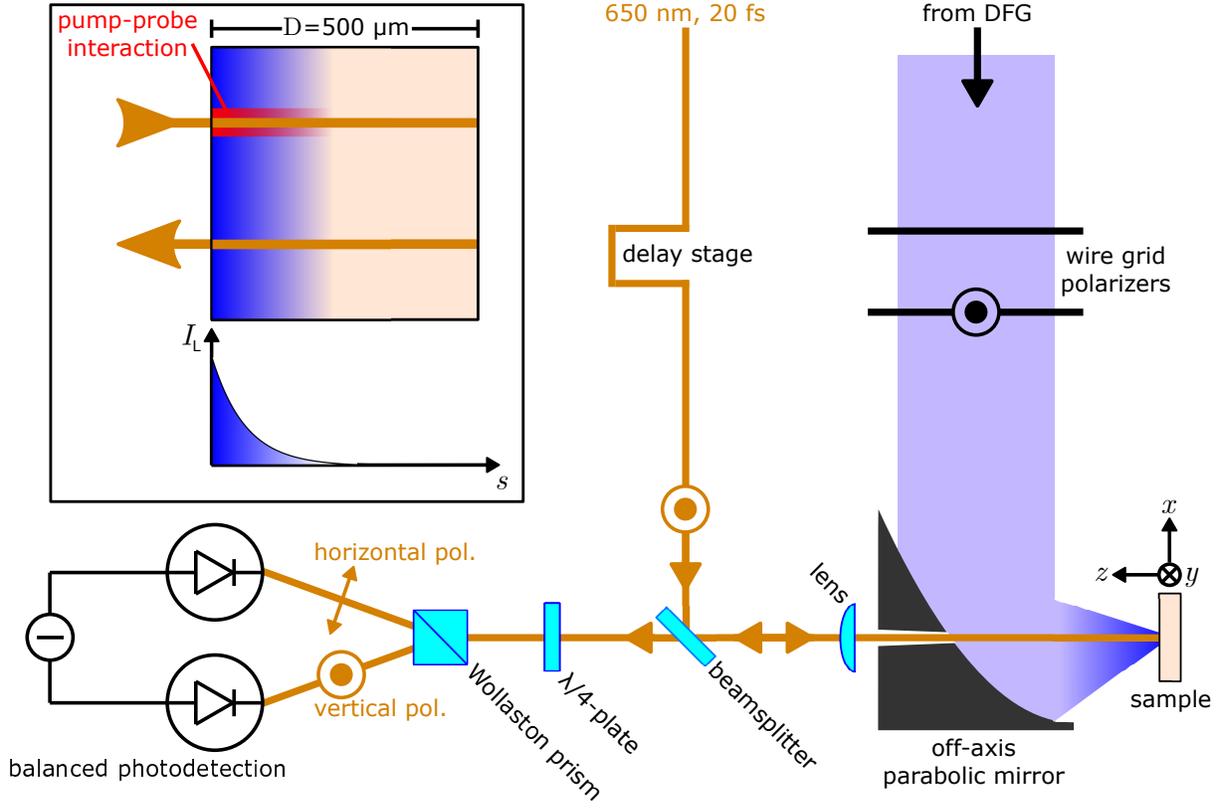}
\caption{\label{fig:exp_setup} Outline of the experimental setup. The pump path is shown in blue, while the probe path is shown in orange. The inset shows an illustration of the probe beam interacting with the pump beam inside the material.}
\end{figure*}

An outline of the experimental setup is presented in Fig.~\ref{fig:exp_setup}. The mid-IR pump radiation coming from a difference frequency generation (DFG) source based on three-wave mixing in GaSe is shown in blue. It propagates through two wire grid polarizers (WGPs) and is focused onto the sample by an off-axis parabolic mirror (OAP). The downstream WGP is fixed at vertical polarization, while the upstream one can be rotated to control the pump fluence \(F\) on the sample. 

The probe beam at 650~nm wavelength propagates over a delay stage and is then reflected off a beamsplitter and focused onto the sample via a trough hole in the OAP. After having been reflected from the sample, the beam is collimated again by the same lens it was focused by and is then transmitted through the beamsplitter. Afterwards, it goes through a \(\lambda\)/4-plate to become circularly polarized for unexcited sample material. Finally, the beam is split into its vertical and horizontal polarization components and then detected as the difference between the voltage levels of equivalent photodiodes. With this layout, the balanced detection is sensitive to changes in the ellipticity of the probe beam.

The inset of Fig.~\ref{fig:exp_setup} illustrates how the intensity of the pump radiation \(I_{\rm L}\) decays when propagating into the sample along \(s\). This decay determines the size of the pump-probe interaction volume (indicated in red), and hence has an influence that is strongly dependent on the exact pump frequency \(\nu_{\rm p}\). After its interaction with the pump beam, the probe is reflected from the sample backside and transmitted through the front surface towards the detection.

\section{Computational details}

For LaAlO$_3$ the point group is D$_{3d}$ which gives two degenerate E$_g$ modes with  Raman tensors of the form
\begin{equation}
    \Pi_1 = \begin{pmatrix} d &\cdot &\cdot \\ \cdot &-d &e \\ \cdot &e &\cdot \end{pmatrix}\label{eq:raman1}
\end{equation}
and
\begin{equation}
    \Pi_2 = \begin{pmatrix} \cdot& -d &-e \\ -d &\cdot &\cdot \\ -e &\cdot &\cdot \end{pmatrix}\label{eq:raman2}.
\end{equation}
In our simulations we assume that the pump electric field is polarized along the x-direction orthogonal to the crystallographic c-axis, and also that we detect only the mode corresponding to $\Pi_1$.  These assumptions imply that only the $xx$ component of $\Pi_1$ is important for estimating the magnitude of ERS.  Note that the exact choice of the electric field polarization within the plane orthogonal to the c-axis has an influence only on the  polarization of the excited E$_g$ mode but not its overall amplitude.

The anharmonic lattice and light-matter interaction potentials, $V_\mathrm{int}$ and $V_\mathrm{lat}$, which capture both ERS and IRS, are then given by
\begin{align}
V_\mathrm{lat} & = \frac{\Omega_\mathrm{IR}^2}{2}Q_\mathrm{IR}^2+\frac{\Omega_\mathrm{R}^2}{2}Q_\mathrm{R}^2+cQ_\mathrm{R}Q_\mathrm{IR}^2, \\
V_\mathrm{int} & = Z_\mathrm{IR} Q_\mathrm{IR} E(t) + \varepsilon_0 d Q_\mathrm{R} E^2(t).
\end{align}
Here, $Q_\mathrm{R}$ and $Q_\mathrm{IR}$ are the phonon amplitudes of the low-frequency Raman-active $E_g$ and high-frequency infrared-active $E_u$ modes, $\Omega_\mathrm{R}$ and $\Omega_\mathrm{IR}$ are the corresponding phonon frequencies, $c$ is the quadratic-linear coupling coefficient, $Z_\mathrm{IR}$ is the mode effective charge, $\varepsilon_0$ is the dielectric constant,  and $E(t)$ is the electric field. 

\begin{figure}
\includegraphics[scale=0.225]{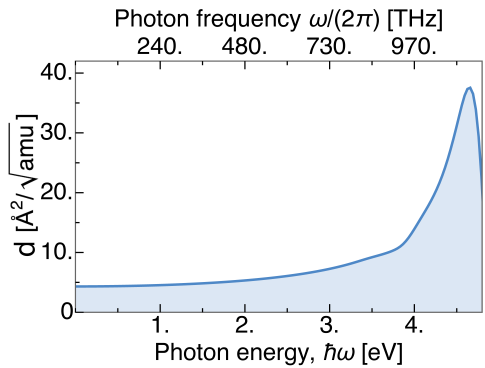}
\caption{\label{fig:Ramantensor} Calculated values of the $d$ component of the frequency-dependent Raman tensor  (see Eqs.~\ref{eq:raman1} and ~\ref{eq:raman2}) over a broad spectral range up to the band gap.}
\end{figure}

We model the electric field component of the pump pulse as 
\begin{equation}
E(t) = E_0 e^{-2 \ln 2 t^2/\tau^2} \cos(\omega_0 t),
\end{equation}
where $E_0$ is the peak electric field, $\omega_0$ is the center frequency, and $\tau$ is the full width at half maximum duration of the pulse. The term proportional to $Q_\mathrm{R}E^2(t)$ describes electronic Raman scattering, and $Q_\mathrm{R}Q_\mathrm{IR}^2$ describes ionic Raman scattering, when $Q_\mathrm{IR}$ is coherently excited through infrared absorption, given by the term $Q_\mathrm{IR} E(t)$. We obtain the phonon amplitudes by solving the coupled equations of motion
\begin{align}
\ddot{Q}_\mathrm{IR} + \gamma_\mathrm{IR}\dot{Q}_\mathrm{IR} + (\Omega_\mathrm{IR}^2+2cQ_\mathrm{R})Q_\mathrm{IR} & = Z_\mathrm{IR} E(t), \label{eq:eom_IR}\\
\ddot{Q}_\mathrm{R} + \gamma_\mathrm{R}\dot{Q}_\mathrm{R} + \Omega_\mathrm{R}^2Q_\mathrm{R} & = cQ_\mathrm{IR}^2(t) + \varepsilon_0 d E^2(t), \label{eq:eom_R}
\end{align}
where $\gamma_\mathrm{IR}$ and $\gamma_\mathrm{R}$ are the phonon linewidths \cite{Juraschek2018a}. As LaAlO$_3$ is a wide band-gap insulator, the component $d$ of the Raman tensor is nearly constant in the mid-IR spectral region, see Fig.~\ref{fig:Ramantensor}.

We calculated the phonon eigenfrequencies, eigenvectors, and the Raman tensors from first-principles using the density functional theory formalism as implemented in the Vienna ab-initio simulation package (VASP) \cite{kresse:1996,kresse2:1996}, and the frozen-phonon method as implemented in the phonopy package \cite{phonopy}. We used the default VASP projector augmented wave (PAW) pseudopotentials for every considered atom and converged the Hellmann-Feynman forces to 25$\times$10$^{-5}$~eV/\AA{} using a plane-wave energy cut-off of 850~eV and a 6$\times$6$\times$6 $k$-point gamma-centered Monkhorst-Pack mesh \cite{Monkhorst/Pack:1976} to sample the Brillouin zone. For the exchange-correlation functional, we chose the PBEsol form of the generalized gradient approximation (GGA) \cite{perdew:2008}. We found that an on-site Coulomb interaction of 12~eV on the La ion reproduces well structural and lattice dynamical properties: our fully relaxed structure with a lattice constant 5.38~\AA{} and a pseudocubic angle of 60.13$^\circ$ fits reasonably well to common experimental values \cite{Berkstresser1991}, as do our calculated phonon frequencies \cite{Willett-Gies2014}. To calculate the frequency-dependent component, $d$, of the Raman tensor we followed the scheme of reference~\cite{porezag:1996}. We computed the quadratic-linear coupling coefficient, $c$, by calculating the total energies as a function of ion displacements along the normal mode coordinates of the $E_g$ and $E_u$ modes and then fitting the resulting three-dimensional energy landscape to 
the potential $V_\mathrm{lat}$. The parameters used in the evaluation are listed in Table~\ref{tab:values}.

\begin{table*}[t]
\centering
\bgroup
\def\arraystretch{1.0}
\caption{
Parameters used in the evaluation of the equations of motion (\ref{eq:eom_IR}) and (\ref{eq:eom_R}). The symbol $\mu$ denotes the atomic mass unit and $e$ the elementary charge.
}

\begin{tabular}{l r l r}
\hline\hline
$\Omega_\mathrm{IR}/(2\pi)^\ast$ & 20.6 THz & ~~~~~$c$ & 20 meV/(\AA{}$\sqrt{\mu}$)$^3$ \\
$\Omega_\mathrm{R}/(2\pi)$ & 1.1 THz & ~~~~~$Z_\mathrm{IR}$ & 0.58 $e$/$\sqrt{\mu}$ \\
$\gamma_\mathrm{IR}^\dagger$ & 0.94 THz & ~~~~~$R(\omega_0\equiv$ mid-IR) & 4.3 \AA{}$^2$/$\sqrt{\mu}$ \\
$\gamma_\mathrm{R}^\ast$ & 0.05 THz & ~~~~~$E_0$ & 12 MV/cm \\
& & ~~~~~$\tau$ & 250 fs \\
\multicolumn{3}{l}{${}^\ast$adjusted to fit experiment, ${}^\dagger$from Ref.~\cite{Willett-Gies2014}} \\
\hline\hline
\end{tabular}
\label{tab:values}
\egroup
\end{table*}

\begin{acknowledgments}
This research was supported by the NCCR MUST, funded by the Swiss National Science Foundation (SNSF).  The project was also funded by the SNSF under project number 200021\_169698 and by the ERC Advanced Grant program, No. 291151. Calculations were performed at the Swiss National Supercomputing Centre (CSCS) supported by the project IDs s624 and p504.  D.M.J. received support from the SNSF under project ID 184259.

\end{acknowledgments}

\bibliography{library_fixed}

\end{document}